\documentclass[12pt]{iopart}
\usepackage{iopams}  
\usepackage{graphicx}
\usepackage{iopams}

\begin{document}

\title[OA Signals in PVA hydrogel phantoms]{Detection and numerical simulation of optoacoustic near- and farfield signals
observed in PVA hydrogel phantoms}

\author{O. Melchert, E. Blumenr\"other, M. Wollweber and B. Roth}

\address{Hanover Centre for Optical Technologies (HOT), Leibniz Universit\"at Hannover, Nienburger Str.\,17, D-30167 Hannover, Germany}
\ead{oliver.melchert@hot.uni-hannover.de}

\begin{abstract}
We present numerical simulations for modelling optoacoustic (OA) signals
observed in PVA hydrogel tissue phantoms.  We review the computational
approach to model the underlying mechanisms, i.e.\ optical absorption of laser
energy and acoustic propagation of mechanical stress, geared towards experiments
that involve absorbing media only. 
We apply the numerical procedure to model signals observed in the acoustic
near- and farfield in both, forward and backward detection mode, in PVA
hydrogel tissue phantoms (i.e.\ an elastic solid).
Further, we illustrate the computational approach by modeling OA signal for 
several experiments on dye solution (i.e.\ a liquid) reported in the 
literature, and benchmark the research code by comparing our fully $3$D 
procedure to limiting cases described in terms of effectively $1$D approaches.
\end{abstract}


\vspace{2pc}


\maketitle


\section{Introduction}
Optoacoustics (OAs) can be considered a two-part phenomenon, consising of two
distinct processes that occur on different time-scales \cite{Kruger:1995}:
fistly, the fast optical absorption of laser radiation (or, more general,
optical radiation) inducing a photothermal heating of the absorbing media, and,
subsequently, the slow emission of ultrasound waves due to thermoelastic
expansion and stress field relaxation.  Albeit thermoelastic expansion is just
one out of several mechanisms that support the conversion of laser radiation to
acoustic stress \cite{Sigrist:1978,Sigrist:1986,Hutchins:1986,Tam:1986}, it is
considered to be the dominant process during experiments similar to those
reported in the presented study.  Whereas the optical absorption is assumed to
occur instantaneously, the acoustic propagation of sound waves is a
comparatively slow process that occurs on a microsecond timescale. 
Assuming the optical absorption to be instantaneous has consequences for the
theoretical treatment of the probelm inherent dynamics
\cite{Diebold:1990,Diebold:1991,Kruger:1995}. It not only allows to decouple
the optical absorption problem from the acoustic propagation problem but also
allows to simplify the latter under the assuptions of \emph{stress confinement}
and \emph{thermal confinement} \cite{Wang:2009}.

Here, we present a combined study, complementing laboratory experiments on
melanin enriched polyvinyl alcohol hydrogel (PVA-H) phantoms via custom
numerical simulations.  The employed PVA-H phantoms represent light absorbing
elastic solids that  are conveniently used to mimic biological soft tissue in
ultrasound experiments and photoacoustic imaging \cite{Xia:2011,Zell:2007}.
With the presented article we build upon a recent study wherein we discussed
detection, numerical simulation and approximate inversion of OA signals
\cite{Blumenroether:2016} on two-layered PVA-H phantoms recorded in the
acoustic far field. Similar to this recent study, we here aim at modeling the
general shape of OA signals resulting from the subtleties of the OA source
volume but make no attempt at modeling the effect of the piezoelectric
transducer used in the laboratory setup as, e.g., done in Ref.\
\cite{Gonzalez:2014}.  In contrast to the above study we here also obtain
signals in the acoustic nearfield, allowing to deduce the optical properties of
the source volume from the measured curves.

The article is organized as follows: in section \ref{sec:theory} we recap the
theoretical framework of OA signal generation, in section \ref{sec:results} we
then summarize our experiments and custom simulations on PVA-H phantoms and in
section \ref{sec:summary} we conclude with a summary. Note that, in order to
benchmark our research code, we replicated several experiments that were
reported in the literature by numerical means in section \ref{sec:appendixA}.

\section{Optoacoustic (OA) signal generation}
\label{sec:theory}

From a computational point of view, the challenge in modelling the OA
phenomenon is greatly reduced by the fact that both processes effectively
decouple and can thus be solved separately.  I.e., once the problem of optical
absorption of laser energy for a given source volume is solved, the result can
be converted to an initial acoustic stress profile for which the acoustic
propagation problem can be solved independently. 

\paragraph{The optical absorption problem:}

In \emph{stress confinement}, describing the limiting case in which the
temporal duration of the irradiation pulse is short enough to be represented
by a delta-function on the scale of typical acoustic propagation times,
photothermal heating can be accounted for by a heating function $H(\vec{r},t)$
that factors in space and time following
\begin{equation}
H(\vec{r},t)=W(\vec{r})\,\delta(t).
\end{equation}
Considering absorbering media (without scattering), a proper Ansatz for the 
volumetric energy density, i.e.\ the amount of laser energy absorbed by the 
medium, reads
\begin{equation}
W(\vec{r})=f_0 f(\vec{r}_\perp)\,\mu_{a}(z)\,\exp\Big\{ -\int_0^z \mu_{a}(z^\prime)~dz^\prime\Big\},\label{eq:BeerLambert}
\end{equation}
where the axial absorption depth profile is in accordance with Beer-Lamberts
law for a depth-dependent absorption coefficient $\mu_{a}(z)$ and where
the factor $f_0$ describes the incident laser fluence.
For the irradiation source profile (ISP) $f(\vec{r}_\perp)$ we here 
consider a flat-top irradiation source profile
\begin{equation}
f(\vec{r}_{\perp}) = 
\cases{ 1, & for $|{\vec{r}}_{\perp}| \leq a$ \\
  \exp\{-(|{\vec{r}}_{\perp}|-a_0)^2/d_0^2\}, & for $|{\vec{r}}_{\perp}|>a$\\}
\label{eq:ISP} 
\end{equation}
where, subsequently, the ISP parameters $a_0$ and $d_0$ are set to be
consistent with beam profiling measurements carried out during our laboratory
experiments. Note that such a beam profile was used previously to compare
numerical simulations with experiments, see, e.g., Ref.\ \cite{Paltauf:1997},
wherein $a_0/d_0\approx4.5$.

In case one might model absorbing and scattering media, the problem of the
interaction of laser radiation with the underlying source volume, the latter
described using its optical properties $\mu_a$ (absorption coefficient),
$\mu_s$ (scattering coefficient), and $g$
(scattering anisotropy parameter) can be solved in terms of a Monte-Carlo
approach of photon migration in media (see Ref.\ \cite{Jacques:2013} for a
thorough review on the optical properties of biological tissue). Therefore,
depending on the inherent ``complexity'' of the source volume configuration,
different publicly available research codes exist. E.g., considering the
problem of steady state light transport in layered media, the translational
invariance of the source volume yields an effective $2$D problem for which
the Greens function response to an infinitely narrow laser beam can be computed
using the {\rm{C}} code {\rm{MCML}} \cite{Wang:1995} (meaning ``Monte Carlo for
Multi-Layered media''; or its {\rm{GPU}} variant
{\rm{CUDAMCML}}, see Ref.\ \cite{Alerstam:2010}).  The full response to a
spatially extended ISP can then be computed using $2$D convolution tools
\cite{Wang:1997,Melchert_FJ:2016}.  A fully ``voxelized'' $3$D representation
of the source volume can be handled using the {\rm{C}} code {\rm{MCXYZ}}
\cite{Jacques_mcxyz:2013}.

\paragraph{The acoustic propagation problem:}
In \emph{thermal confinement}, describing the limiting case in which the laser
pulse duration is significantly shorter than the thermal relaxation time within
the source volume, the propagation of initial acoustic stress profiles $p_{\rm
0}(\vec{r})=\Gamma W_{\rm 0}(\vec{r})$ is governed by the scalar $3$D OA wave
equation
\begin{equation}
\big[ \partial_t^2 - c^{2} \Delta \big]~p(\vec{r},t) = 
        \partial_t~p_0(\vec{r})\,\delta(t), 
        \label{eq:OAWaveEq}
\end{equation}
which yields the excess pressure field $p(\vec{r},t)$ at time $t$ and field
point $\vec{r}$ \cite{Gusev:1993,Wang:2009}.
Above, the Gr\"uneisen parameter $\Gamma$ represents an effective material 
parameter that characterizes the efficiency of the conversion of deposited
laser energy to actual acoustic stress. For a recent discussion of the 
intricate procedure of determining $\Gamma$ for biological media using
photoacoustic spectroscopy see Ref.\ \cite{Yao:2014b}.

Assuming an acoustically homogeneous medium,
a closed form solution for the excess pressure profile 
$p(\vec{r},t)$ as function of the observation time $t$ at the field
point $\vec{r}$ is possible in terms of the OA Poisson integral
\begin{equation}
p(\vec{r},t) = \frac{\Gamma}{4\pi c} \partial_t
\int\limits_{V}\!\frac{W(\vec{r}^\prime)}{|\vec{r}-\vec{r}^\prime|}
\delta(|\vec{r}-\vec{r}^\prime| - ct)\,\mathrm{d}\vec{r}^\prime
,\label{eq:PAPoissonEq}
\end{equation} 
where the source volume $V$ signifies the part of the computational domain
wherein $W(\vec{r})\neq 0$, and $\delta(\cdot)$ limiting the
integration to a time-dependent surface constraint by $|\vec{r}-\vec{r}^\prime|
= ct$.  

\paragraph{Signal detection using a thin piezoelectric transducer:}
In order to detect OA signals we here use a piezoelectric polyvinylidenfluoride
(PVDF) transducer foil with a thickness of approximately $10\,{\rm{\mu m}}$ 
and a circular active area with $1\,{\rm mm}$ diameter.
Without elaborating on the subtleties of piezoelectric signal conversion via
PVDF transducers, note that the voltage response of the electrical circuit
consisting of the transducer plus a signal amplifier is proportional to the
average acoustic stress inside the transducer foil
\cite{Schoeffmann:1988,Paltauf:1997}, i.e.\ $U(t)\propto \bar{p}(t)$. However,
if the typical extension of the acoustic wave transmitted by the underlying
medium is large compared to the thickness of the transducer foil, the voltage
response is simply proportional to the pressure averaged over the foil surface
\cite{Sigrist:1978,Jaeger:2005}, i.e.\ $U(t)\propto \int_A p(\vec{r},t)\,\mathrm{d}A$. Since
in our experiments absorbing layers within the medium have a thickness of
approximately $1\,{\rm mm}$, supporting acoustic waves with a spatial extend
that exceeds the transducer foil thickness by a factor of approximately $100$,
we might compare the measured OA signals to computer simulations based on
solving Eq.\ (\ref{eq:PAPoissonEq}) for a field point $\vec{r}$, signifying the
location of the transducer surface. In some cases, as, e.g.\ in numerical
experiment number three in \ref{sec:appendixA}, it might also be necessary to
integrate the resulting (singular) pressure signals over a set of points
representing the transducer surface.

\paragraph{Phenomenology of OA signals:}

\begin{figure}[t!]
\centerline{\includegraphics{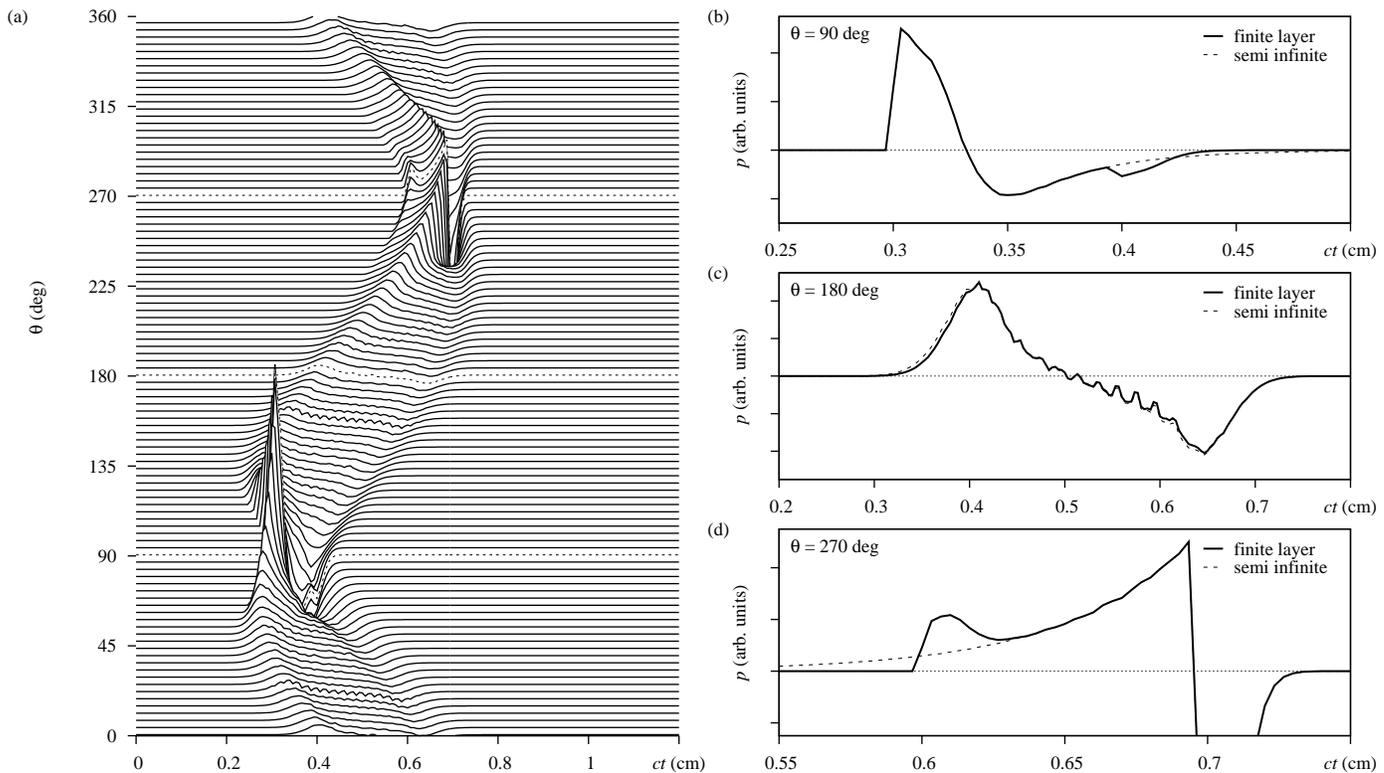}} 
\caption{Numerical experiments to model OA signals resulting from the
absorption of laser light by a homogeneously absorbing layer.
(a) stacked plot of signals resulting from measurements via 100 point detectors 
arranged as a circular array enclosing the region of interest. The distinguished
signals at $\theta=90^{\circ}$, $180^{\circ}$, and $270^{\circ}$ (indicated as dashed lines) are 
shown in more detail in the subsequent subplots.
(b) backward mode signal ($\theta=90^{\circ}$) recorded in the early farfield at $D=1.11$,
(c) side-view recorded at $\theta=180^{\circ}$, and,
(d) forward mode signal ($\theta=270^{\circ}$) recorded in the farfield at $D=2.22$.}
\label{fig:theory}
\end{figure}

During the course of their propagation, the initial acoustic stress profiles
$p_{\rm 0}(\vec{r})$ experience a shape change induced by diffraction,
dispersion, acoustic attenuation and nonlinear acoustic effects
\cite{Sigrist:1986}.  In the remainder we restrict our discussion to the shape
transformation due to diffraction, which, for the considered laser intensities 
and material properties is assumed to be dominant.
The characteristic features of OA signals are sensitive to the geometric
subtleties of the ISP and on the precise detection setup, i.e.\ the
detector-to-sample distance and the geometry and spatial extend of the
detection device.
In principle one can distinguish two main detection regimes: the acoustic
nearfield, wherein an initially plane acoustic wavefront is considered to be 
of plane wave type, and the acoustic farfield, wherein the wavefront is 
considered to effectively assume a spherical shape.
Both might be distinguished using the dimensionless diffraction parameter 
\begin{equation}
D=\frac{|z_{\rm D}| \lambda_{\rm{ac}}}{r_{\rm 0}^2}, \label{eq:diffpar}
\end{equation}
with detector-to-sample
distance $|z_{\rm{D}}|$, beam diameter $r_0$ (here interpreted as the
$1/e$-intensity threshold of the flat-top ISP in Eq.\ (\ref{eq:ISP}), i.e.\
subsequently $r_0 \equiv a_0+d_0$) and characteristic acoustic wavelength
$\lambda_{\rm{ac}}$ \cite{Sigrist:1986}, which, in stress confinement 
reads $\lambda_{\rm{ac}}=2\, \mu_{\rm{eff}}^{-1}$.  
Subsequently, we consider a single absorbing layer in between optically
transparent layers, thus we here assume the effective attenuation coefficient
$\mu_{\rm{eff}}\equiv \mu_{a}$.
Near- and farfield conditions are realized for $D<1$ and $D>1$, respectively. 

So as to facilitate intuition regarding the generic features of OA signals,
Fig.\ \ref{fig:theory} illustrates a sequence of simulated excess pressure
curves. The simulation setup reads as follows: 
(i) as \emph{source volume} we consider a $3$D mesh with side lengths
$(L_x,L_y,L_z)=(1\,{\rm cm},1\,{\rm cm},1\,{\rm cm})$ and
$(N_x,N_y,N_z)=(900,900,300)$ mesh points. Therein, a single absorbing layer
(tangent vectors $\vec{e}_x$ and $\vec{e}_y$) with homogeneous absorption
coefficient $\mu_a=24\,{\rm cm}^{-1}$ extends from $z=0 - 0.1\,{\rm cm}$.
(ii) as \emph{irradiation source} we consider a flat-top laser beam as
described by Eq.\ (\ref{eq:ISP}) with beam parameters $a_0=0.1\,{\rm cm}$ and
$d_0=a_0/2$, i.e.\ $1/e$-intensity threshold $r_0=0.15\,{\rm cm}$.  The
symmetry axis of the beam is considered to coincide with the $z$-axis, defining
a plain normal irradiation scenario.
(iii) an array of $100$ \emph{point detectors} is arranged along a circle with
center $(x_{\rm c},y_{\rm c},z_{\rm c})=(L_x/2,L_y/2,0.2\,{\rm cm})$ and
radius $R=0.5\,{\rm cm}$ in the $xz$-plane, surrounding the region of the
source volume with nonzero $p_0(\vec{r})$. The individual detector
positions are generated following $(x_{\rm D},y_{\rm D},z_{\rm D})=(x_{\rm c}+R
\cos(\theta),L_y/2,z_{\rm c}-R \sin(\theta))$ with $\theta$ in increments of 
$2\pi/100$.
For convenience, since we are only interested in the generic shape of the
waveforms as function of $\theta$, we set $c=0.15\,{\rm cm/\mu s}$,
$f_0=1\,{\rm J/cm^2}$ and $\Gamma=1$.

Fig.\ \ref{fig:theory}(a) shows a stacked representation of the sequence of
measurements recorded using the circularly arranged point detectors. As evident
from the figure, both, the signal amplitude and the waveform change as function
of $\theta$. The distinct signal shapes can be explained by the relative
orientation of the detection point and Beer-Lambert absorption profile within
the source volume. The three measurement angles $\theta=90^{\circ}$ (backward mode),
$\theta=180^{\circ}$ (side-view), and $\theta=270^{\circ}$ (forward mode), distinguished as
dashed lines in the plot, are detailed in Figs.\ \ref{fig:theory}(b-d).  In
backward mode the detection point is located at $D=1.11$, thus observing a
borderline farfield signal (solid line).  As can be seen from Fig.\
\ref{fig:theory}(b) it consists of an initial compression peak, indicating the
beginning of the absorbing layer, followed by an extended rarefaction phase and
a pronounced rarefaction dip, marking the end of the absorbing layer as evident
from a comparison to the simulation result for a semi-infinite layer (indicated
by the dashed line in Fig.\ \ref{fig:theory}(b)). These are the characteristic
features that allow for OA depth profiling in the acoustic far field. Further
exemplary backward mode signals in the acoustic near and farfield can be found
in numerical experiments $1$ and $2$ detailed in \ref{sec:appendixA}. In the
side-view at $\theta=180^{\circ}$ the numerical procedure yields a quite symmetric
bipolar signal, see Fig.\ \ref{fig:theory}(c), and in forward mode at
$\theta=270^{\circ}$ a sequence of compression peaks terminated by a strong rarefaction
phase is visible, see Fig.\ \ref{fig:theory}(d). A further exemplary forward
mode signal can be found in numerical experiment $3$ in \ref{sec:appendixA}.
Note that in this case, the initial compression peak is due to the finite
extend of the absorbing layer and the signal is observed in the farfield at
$D=2.22$.  Further, note that while the signal features in $\theta=90^{\circ}$ and
$270^{\circ}$ reflect the width of the absorbing layer, the side-view at $\theta=180^{\circ}$
relates to the width of the illuminated region.

Subsequently, we report on both, laboratory experiments and numerical
simulations that were carried out to better understand the features of OA
signals resulting from the absorption of laser energy by melanin enriched
elastic solids. 
Following the formal decomposition of the problem of OA signal generation, the
numerical simulation were performed by (i) modelling the optical properties of
the computational domain guided by the experimental setup, (ii) considering the
absorption of laser light within the source volume, and, (iii) solving the
stress wave propagation problem for a medium with homogeneous acoustic
properties.  In order to benchmark our research code we first replicated
several laboratory experiments on liquid dye solution reported in the
literature and illustrated in \ref{sec:appendixA}.
Note that while there are further numerical approaches that yield OA signals
and their diffraction characteristics in terms of a simplified $1$D approach
\cite{Terzic:1984,Karabutov:1996}, see \ref{sec:appendixB}, we prefer the
presented voxelized $3$D representation of the source volume since it allows to
model custom beam profiles and also more intricate absorbing structures.
Further, note that a similar approach for coupling the optical deposition and
acoustic propagation problem has been discussed by Ref.\ \cite{Jacques:2014}.

\section{OA signals in PVA hydrogel phantoms}
\label{sec:results}

As reviewed recently in Ref.\ \cite{Fonseca:2016}, ideal phantoms for
photoacoustic imaging should meet several criteria in order to ascertain
tissue-realistic photoacoustic properties and fabrication reproducibility.
Here, in our combined experimental and numerical study we consider polyvinyl
alcohol based hydrogel (PVA-H) phantoms \cite{Kharine:2003,Wollweber:2014}.
Albeit the preparation procedure of the PVA-H phantoms is rather extensive, as
evident from the detailed preparation protocol to yield melanin enriched tissue
phantoms summarized in Ref.\ \cite{Blumenroether:2016}, we appreciate their
structural rigidity and tissue-like acoustic properties, rendering them a
suitable tool to mimic soft tissue in ultrasound experiments \cite{Zell:2007}
and photoacoustic imaging \cite{Xia:2011}. 

In the remainder we complement OA signals measured in PVA-H tissue phantoms
consisting of a single, melanin enriched absorbing layer in between optically
transparent layers, with numerical calculations resulting from the two-part
modelling approach discussed in Section \ref{sec:theory}.  In our experiments,
laser pulses were generated using an optical parametric oscillator (NL303G +
PG122UV, Ekspla, Lithuania) at a wavelength of $\lambda=532\,{\rm{nm}}$ with
pulse duration of approximately $6\,{\rm{ns}}$. After traversing a fiber with
$800\,{\rm{\mu m}}$ core diameter (Ceramoptec, Optran WF 800/880N), the beam
profile was found to be well described by a flat-top shape, consistent with
Eq.\ (\ref{eq:ISP}).  For the detection of OA pressure signals, a piezoelectric
transducer, consisting of a $9\,{\rm{\mu m}}$ thick piezoelectric
polyvinylidenfluorid (PVDF) film with approximately $50\,{\rm{nm}}$ indium tin
oxide (ITO) electrodes and an circular active area with radius $r_{\rm
D}=0.5\,{\rm{mm}}$, is used \cite{Niederhauser:2005}.

\paragraph{Experiment 1:}

\begin{figure}[t!]
\centerline{\includegraphics{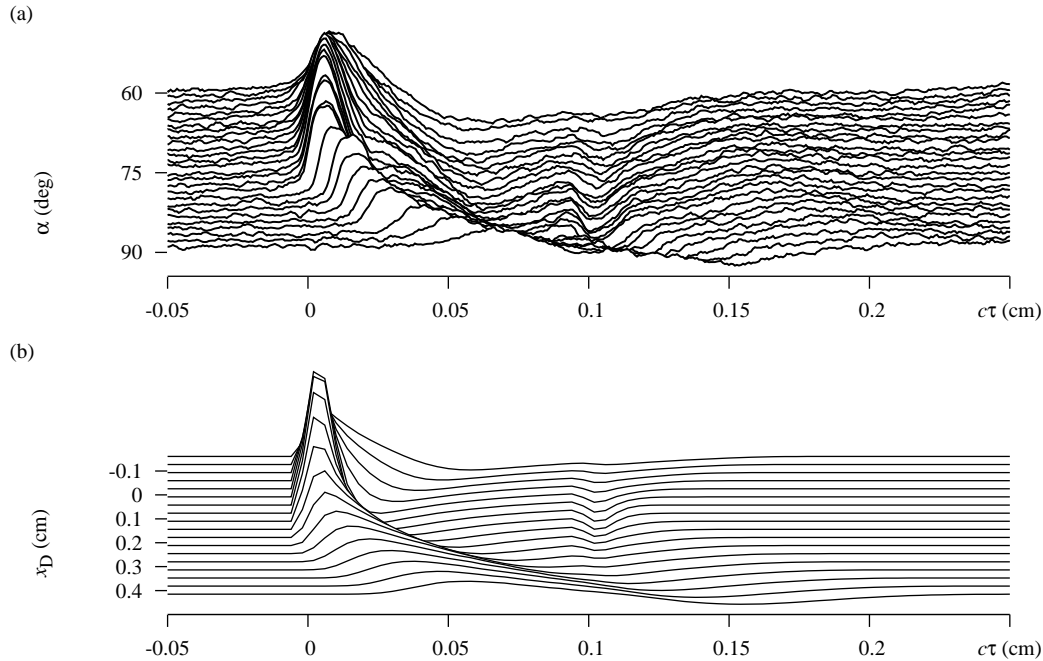}} 
\caption{Results of a qualitative study involving laboratory experiments on
PVA-H phantoms and custom numerical simulations to predict the expected OA
signals.  Both figures show normalized OA pressure profiles $p$ (consising of a
pronounced compression peak, an extended rarefaction phase and a pronounced
rarefaction dip) as function of the retarded signal depth $c \tau$.  (a) OA
signals measured on a melanin enriched PVA-H phantom with setup paramters
reported in the text. Upon decreasing the parameter $\alpha$, the laser spot is
shifted across the phantom towards the detector location.  (b) OA signals
obtained via numerical experiments, modeling the laboratory setup with the
parameters reported in the text. Considering plane normal irradiation, $x_{\rm
D}$ indicates the deviation of the detector position from the symmetry axis of
the beam. }
\label{fig:ex1}
\end{figure}

First, we present a qualitative comparative study where we complement 
measurements, in which a (pulsed) laser beam was scanned across a PVA-H phantom while
keeping the detector position fixed, with custom numerical simulations. 
Albeit the precise implementation of the numerical simulation differs slightly
from the laboratory setup, the resulting OA signals can nevertheless be compared
on a qualitative basis.
To be more precise, in the laboratory setup, the fiber transmitting the laser
pulse was hinged above the specimen holder containing the PVA-H phantom. By
pivoting the transmitting fiber in an angular range of $\alpha = 90$ (plane
normal irradiation) through $60$ (inclined irradiation) the illuminated region
was shifted across the phantom. In any case, the laserlight entered
the phantom close to the detector. To facilitate intuition on the measured OA
signals shown in Fig.\ \ref{fig:ex1}(a), note that the illuminated region
approaches the detector as $\alpha$ decreases.
During our experiment, the laser spot diameter on the sample surface was
approximately $0.3\,{\rm cm}$, defining the lengthscale for the
$1/e$-intensity radius for our subsequent numerical simulations. The
measurement was performed in backward mode with a distance $z_{\rm
D}=1\,{\rm{cm}}$ of the absorbing layer to the detector.  The absorbing layer
was prepared with thickness $d_z=0.1{\rm{cm}}$ and absorption coefficient in
the range $\mu_a= 20-24\,{\rm cm}^{-1}$.  I.e.\ the measurement is performed
under farfield conditions at $D=4.4$ (assuming $\mu_a=20\,{\rm{cm}}^{-1}$).  In
Fig.\ \ref{fig:ex1}(a), the individual signals are shown as function of the
time-retarded signal depth $c \tau = c t - z_{\rm D}$ with $c
\tau=0$ signifying the start of the absorbing layer. The speed of sound within
the PVA-H phantom is assumed to be $c=0.15\,{\rm{cm/\mu s}}$.
In the range $\alpha= 60 \ldots 75$ the initial compression peaks appear at a
similar level, indicating the plane part of the acoustic wavefront, resulting
from the high-intensity part of the laser spot.  In the range $\alpha=75 \ldots
90$, a spherical bending of the wavefront due to diffraction can be observed,
causing the excess pressure peaks to reach the detector position with some
delay, manifested as a peak position at $c \tau > 0$.
Note that for an extended range of $\alpha$, a pronounced rarefaction dip at 
$c \tau \approx 0.1\,{\rm{cm}}$ indicates the end of the absorbing layer.

\begin{figure}[t!]
\centerline{\includegraphics{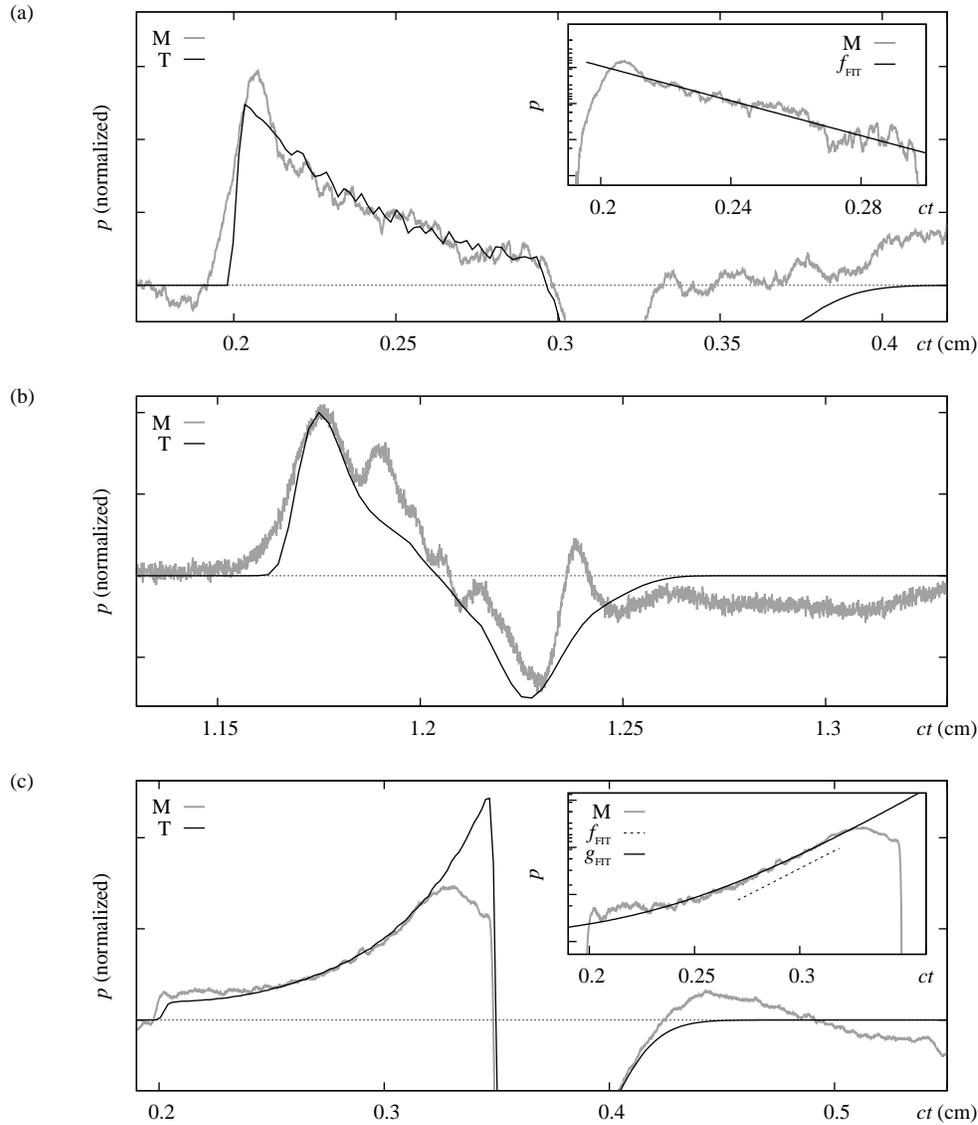}} 
\caption{Measurement and numerical simulation of OA signals resulting from the
absorption of laser light by a PVA-H phantoms.  (a) The main plot shows
measured (labeled ``M'') and computed (labeled ``T'') nearfield signals
recorded in backward mode and the inset illustrates a fit using a simple
exponential function as discussed in the text,
(b) measured and computed farfield signals obtained for a side-view configuration,
(c) the main plot shows measured and computed nearfield signals and the inset
illustrates a fit using two different exponential fitting functions discussed
in the text.}

\label{fig:ex2}
\end{figure}

For our custom numerical simulations shown in Fig.\ \ref{fig:ex1}(b) 
we consider a pointlike detector (initial numerical experiments suggested that
for the given farfield configuration, the spatial extend of the detector 
can be neglected), a flat-top ISP with beam parameters 
$a_0=0.15\,{\rm{cm}}$ and $a_0/d_0=2$ and a homogeneous absorption coefficient
of $\mu_a=20\,{\rm{cm}}^{-1}$ across the $0.1\,{\rm{cm}}$ thick absorbing 
layer. 
Further, in contrast to the experimental setup, plain normal laser irradiation
was assumed with $x_{\rm D}$ (see Fig.\ \ref{fig:ex1}(b)) describing the
deviation of the detector position from the symmetry axis of the beam.
As evident from Fig.\ \ref{fig:ex1}, the OA signals of the laboratory
experiments and numerical simulations agree well on a qualitative basis.

\paragraph{Experiment 2:} In Fig.\ \ref{fig:ex2}(a) we illustrate a comparison
of both, measurement and numerical simulation, for a nearfield signal recorded
in backward mode. To summarize the experiment: we considered an absorbing layer
of thickness $0.1\,{\rm{cm}}$ and absorption coefficient
$\mu_a=20-24\,{\rm{cm}}^{-1}$, the diameter of the laser spot was estimated as
$0.4\,{\rm{cm}}$ and the detector was positioned at a distance of $0.2\,{\rm
cm}$ from the absorbing layer.  As can be seen from the figure, the initial
compression phase in the signal range $ct=0.2 - 0.3\,{\rm{cm}}$ follows a
Beer-Lambert decay, see Eq.\ \ref{eq:BeerLambert}, and experiment and
simulation agree well.  However, note that the geometry of the diffraction
induced rarefaction phase is not well represented by the strict flat-top ISP
modeled using the parameters $a_0=0.21\,{\rm{cm}}$ and $a_0/b_0=2$.  As shown
in the inset of Fig.\ \ref{fig:ex2}(a), anticipating a Beer-Lambert decay and
fitting an exponential function $f_{\rm{FIT}}(x)=a \exp[-\mu_{a}^\prime
c(t-t_0)]$ to the data in the range $ct\in(0.21,0.28)$, yields the fit
parameters $a=O(1)$, $\mu_{a}^\prime=22.1(2) \,{\rm{cm}}^{-1}$ and $ct_0\approx
0.20$ (the latter reflecting that we here show OA signals where the initial
compression peak is normalized to unit height).

\paragraph{Experiment 3:} In Fig.\ \ref{fig:ex2}(b) we show a custom simulation
for an OA signal measured in a side-view configuration (similar to Fig.\
\ref{fig:theory}(c)). The measurement was performed at an approximate distance
of $1.7\,{\rm{cm}}$ to the narrow laser spot which had a diameter of
approximately $0.05\,{\rm{cm}}$. For the numerical experiments we used a
flat-top ISP with beam parameters $a_0=0.022\,{\rm{cm}}$ and $a_0/b_0=2$. This
measurement was taken out of a sequence of experiments on tissue phantoms with
a somewhat lower absorption coefficient of $\mu_a \approx 1\,{\rm cm}^{-1}$.  
Note that while both, the theoretical and experimental signal exhibit a 
bipolar shape, the latter appers to be more rugged. This effect is most likely 
due to signal reflections within the PVA-H backing layer used to terminate the 
PVDF transducer.

\paragraph{Experiment 4:}
Finally, in Fig.\ \ref{fig:ex2}(c) we show an OA signal recorded in forward
mode, where the detector is located at a distance of $0.35\,{\rm cm}$ below the
beginning of the absorbing layer. The layer itself has a thickness of
$0.15\,{\rm{cm}}$, i.e.\ $z_D=0.2\,{\rm{cm}}$, 
and an absorption coefficient of $\mu_a=20-24\,{\rm cm}^{-1}$.
During the experiment, the laser spot had a diameter of approximately
$0.2\,{\rm cm}$. For the numerical experiments we used the flat-top beam
parameters $a_0=0.1\,{\rm{cm}}$ and $a_0/b_0=2$ as well as
$\mu_a=24\,{\rm{cm}^{-1}}$. I.e., simulations are performed in the acoustic 
near field at $D=0.74$  The steep increase of the pressure signal in the range
$ct \in (0.2\,{\rm cm},0.35\,{\rm cm})$ is in agreement with a Beer-Lambert law
as seen from the detector position in forward mode. The finite size of the
laser beam is responsible for the strong rarefaction phase in the range $ct \in
(0.35\,{\rm cm},0.45\,{\rm cm})$. In principle, experiment and simulation agree
quite well. Note that, as in experiment 1 above, it is possible to extract the
absorption coefficient from the measured data.  Considering a simple
exponential increase of the pressure signal in the compression range already
yields $\mu_a^\prime=21.0(1)\,{\rm cm}^{-1}$. An effective fitting function of
the form $g_{\rm{FIT}}=a_0 + a_1 \exp[\mu_a^\prime c (t-t0)]$, allowing to also
account for an initial ``flat'' part of the compression phase yields the fit
parameter $a_0=O(10^{-1})$, $a_1=O(1)$, $ct_0\approx 0.36$, and
$\mu_a^\prime=23.7(2)\,{\rm cm}^{-1}$, in reasonable agreement with the
expected range of $\mu_a$.


\section{Summary and Conclusions}
\label{sec:summary}

We presented a combined study, complementing measured OA signals with custom
numerical simulations. The laboratory experiments where performed on melanin
enriched PVA-H phantoms, i.e.\ elastic solids that possess tissue-like acoustic
properties and are used to mimic soft tissue in ultrasound experiments and
photoacoustic imaging \cite{Zell:2007,Xia:2011}.  Since the experimental
conditions satisfy the limits of stress confinement, the problem of OA signal
generation decouples into the processes of laser absorption and stress wave
propagation that can be approached separately.  Consequently, numerical simulations
were performed by: (i) modelling the optical properties of the computational
domain, (ii) considering the absorption of laser light within the source
volume, and, (iii) solving the stress wave propagation problem for a medium
with homogeneous acoustic properties using the optoacoustic Poisson integral.
So as to benchmark our research code we first replicated several laboratory
experiments on liquid dye solution reported in the literature, see
\ref{sec:appendixA}.
Complementing a recent study wherein we elaborated on the detection, numerical
simulation and approximate inversion of OA signals \cite{Blumenroether:2016},
we here apply the procedure to model OA near- and farfield signals observed in
experiments on PVA-H phantoms. Overall we found a good qualitative agreement
between both. In the acoustic nearfield we found that the initial excess
pressure profile could be modeled well using the parameters of the respective
experiments, however, the precise shape of the subsequent rarefaction phase
differed from that recorded in the experiment, see Figs.\ \ref{fig:ex2}(a,c). 
In this regard, note that here the focus is on modelling the principal shape of the OA signal.
We did, however, make no attempt at modelling the behavior of the 
PVDF-based transducer used in the laboratory setup (as previousely done for 
a similar detection device, see Ref.\ \cite{Gonzalez:2014}).
Further, note that a fit of simple model functions to the compression phases in
experiments 2 and 4, see Figs.\ \ref{fig:ex2}(a,c), allowed to obtain estimates
of the absorption coefficients for the absorbing layers from the measurement.
The resulting estimates are in good agreement with the values expected on basis
of the phantom preparation procedure.

\ack{
O.\ M.\ acknowledges support from the VolkswagenStiftung within the
``Nieders\"achsisches Vorab'' program in the framework of the project ``Hybrid
Numerical Optics''  (Grant ZN 3061).  E.\ B.\ acknowledges support from the
German Federal Ministry of Education and Research (BMBF) in the framework of
the project MeDiOO (Grant FKZ 03V0826).  Further valuable discussions within
the collaboration of projects MeDiOO and HYMNOS at HOT are gratefully
acknowledged.
The simulations where performed using {\rm PyPCPI} \cite{Melchert_PyPCPI:2016}, a software tool for 
optoacoustic signal generation developed at the 
Hanover Centre for Optical Technologies.}

\appendix
\section{Numerical simulation of OA signals measured on dye solution}
\label{sec:appendixA}

Subsequently we illustrate the numerical calculation of OA signals that result
from the absorption of laser light in liquid samples, consisting of a layer of
light absorbing dye solution in between layers of optically transparent liquid,
for three different experimental setups reported in the literature.  We
replicated these studies by numerical means for benchmarking our research code.

\begin{figure}[t!]
\centerline{\includegraphics{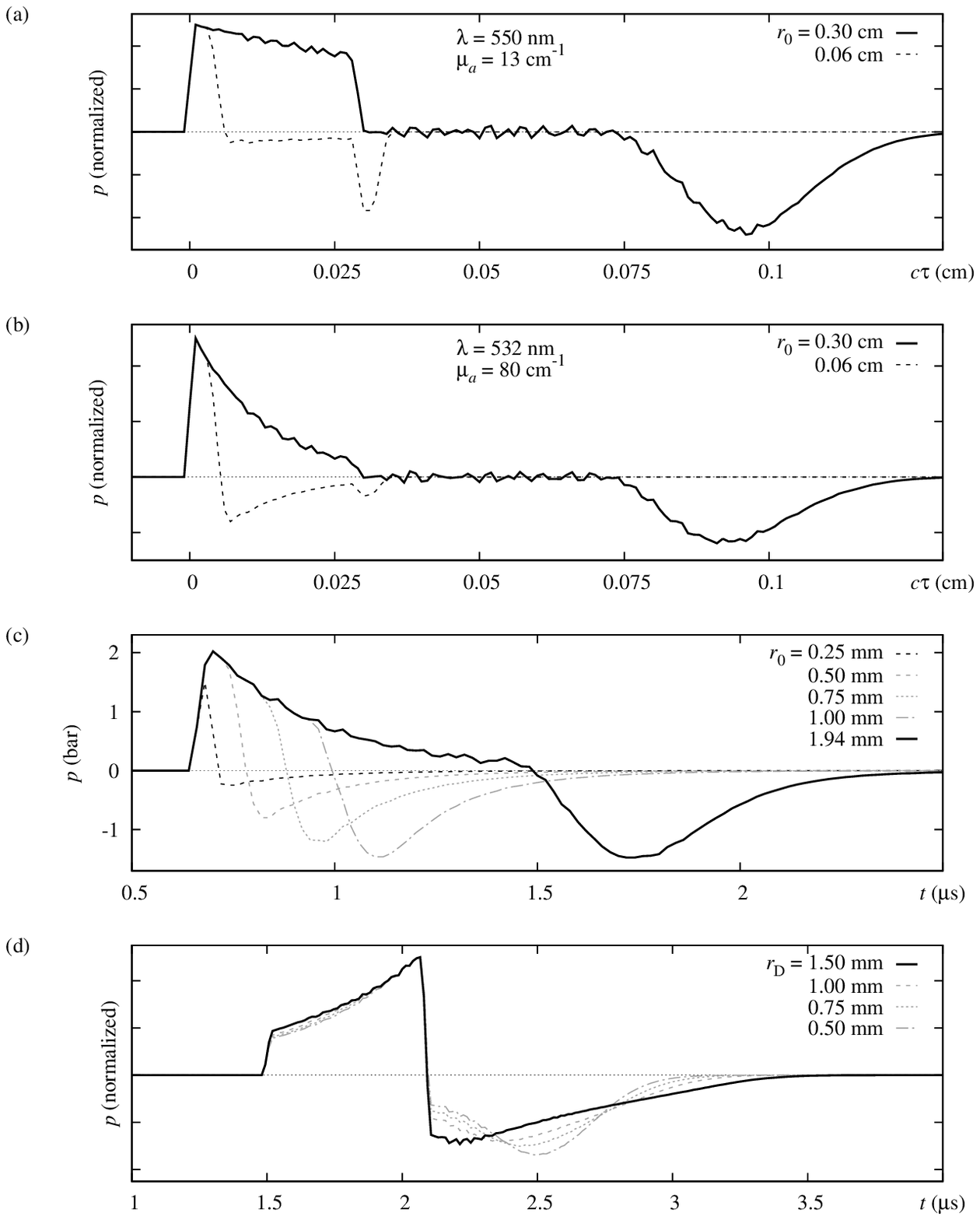}} 
\caption{Numerical experiments to model OA signals resulting from the
absorption of laser light in liquid samples, consisting of a layer of dye
solution in between layers of optically transparent liquid. The subfigures
refer to the laboratory experiments reported in (a) Fig.\ 8(a) of ref.\
\cite{Paltauf:2000}, (b) Fig.\ 8(b) of ref.\ \cite{Paltauf:2000}, (c) Fig.\
3(a) of ref.\ \cite{Paltauf:1996} and Figs.\ 5 and 6 of ref.\
\cite{Paltauf:1997}, (d) Fig.\ 2(b) of ref.\ \cite{Jaeger:2005}. Details on the
setup of the simulation domain can be found in \ref{sec:appendixA}.}
\label{fig:useCase}
\end{figure}

\paragraph{Numerical experiment 1:}
In Ref.\ \cite{Paltauf:2000}, the authors report on an experimental study,
where they observed near and far field signals using an optical detector for
the measurement of acoustic stress waves. The detector was operated in
reflection mode, similar to the backward mode used in our study.  Keeping the
sample to detector distance constant and tuning the width of the flat-top
profile of the HeNe laser used in their setup, they were able to record OA
signals in both measurement regimes.  In Figs.\ \ref{fig:useCase}(a,b) we
illustrate numerical simulations geared towards the laboratory experiments
reported in Ref.\ \cite{Paltauf:2000}.  Therefore, we considered a
computational domain modeling an absorbing layer of width $w=300\,\mu{\rm m}$
and absorption coefficient $\mu_a=13\,\rm{cm}^{-1}$ (i.e.\ weakly absorbing
scenario at $\lambda=550\,{\rm nm}$; see Fig.\ \ref{fig:useCase}(a)) and
$\mu_a=80\,\rm{cm}^{-1}$ (i.e.\ strongly absorbing scenario at
$\lambda=532\,{\rm nm}$; see Fig.\ \ref{fig:useCase}(b)) for diffraction
parameters $D=0.16$ and $D=4$ at a sample to detector distance of $|z_{\rm
D}|=1.2\,{\rm mm}$, referring to the acoustic near field and far field of the
measurement setup, respectively.  Note that the small peak at the signal depth
$\approx 600\,{\rm \mu m}$ in Fig.\ 8b of Ref.\ \cite{Paltauf:2000}, resulting
from acoustic reflection at the plastic container walls that separate the
distinct liquid layeres, is not modeled by our approach. Further, note that the
above laboratory experiments where reported using normalized optoacoustic
signals wherein the initial excess pressure peak has unit height.  Hence, the
numerical simulations tailored towards those experiments are normalized in the
same manner to facilitate a compoarison of both.

\paragraph{Numerical experiment 2:}
In contrast to the previous numerical experiment, where the reference curves
consist of normalized excess pressure signals, a further study reported in
Refs.\ \cite{Paltauf:1996,Paltauf:1997} allows to compare a custom numerical
simulation to an actual pressure signal recorded in the acoustic near field in
reflection mode. The corresponding measurement setup reported in the above
references consists of an (calibrated) optical stress detector and a Nd:YAG
laser at wavelength $\lambda=532 \rm{nm}$ with $H_0=0.18\,{\rm J/m}$ and
flat-top ISP parameters $a_0=1.94\,{\rm mm}$ and
$d_0=0.43\,{\rm mm}$, incident onto a layer of dye solution characterized by a
Gr\"uneisen paremeter $\Gamma=0.11$ and absorption coefficient $\mu_a=24\,{\rm
cm}^{-1}$ at a sample to detector distance of $|z_{\rm D}|=1\,{\rm mm}$ (see Fig.\ 5 of
Ref.\ \cite{Paltauf:1996}).  Further, Ref.\ \cite{Paltauf:1996} reports several
normalized OA signals resulting from different values for the diameter of the
flat-top laser profile to illustrate the change in signal shape during the
transition from the acoustic near field to the far field (see Fig.\ 6 of Ref.\
\cite{Paltauf:1996}).  Our numerical simulations geared towards those
experimets are shown in Fig.\ \ref{fig:useCase}(c).

\paragraph{Numerical experiment 3:}
As a further challenge we considered the laboratory experiments reported in
Ref.\ \cite{Jaeger:2005}. Therein, two sets of experiments with varying
detector and beam-width combinations where reported. Among those, a
piezoelectronic ITO transducer (similar to the one used in our experimental
setup) was used to record OA signals in forward mode.  In detail, 
our numerical simulations consider a circular ITO detector with diameter of $r_{\rm D}=3\,{\rm mm}$ placed at a
distance of $z_{\rm D}=2.25\,{\rm mm}$ of a dye layer with
width $w=0.9 \,{\rm mm}$ and absorption coefficient $\mu_a=1.65\,{\rm
mm}^{-1}$.  As irradiation source, the laboratory setup in Ref.\ \cite{Jaeger:2005} 
used a Nd:YAG laser (operating at $\lambda = 577\,{\rm nm}$) to obtain a
homogeneous circular ISP with a radius of $r_0=3\,{\rm mm}$.  For this
particular setup, the resulting OA signal is strongly distorted by diffraction,
causing an extended rarefaction phase subsequent to the initial raise of the
excess pressure profile expected by Beers law (see Fig.\ 2 of Ref.\
\cite{Jaeger:2005}). In order to model these signal features properly, the
spatial extend of the detector has to be taken into account via integrating the
excess pressure signal numerically over an appropriate circular area. A
sequence of numerical simulations that illustrate the corresponding OA signal
for different detector radii is illustrated in Fig.\ \ref{fig:useCase}(d).
Note that, while Ref.\ \cite{Jaeger:2005} quotes a homogeneous circular beam
profile with $r_0=3\,{\rm mm}$, corresponding to a sharp flat-top ISP, we
observed that a more realistic and smooth flat-top ISP with $1/e$-intensity
radius $r_0$ and beam profile parameters $r_0=a_0+d_0$, where $a_0/d_0=4$,
reproduces the curves observed by Jaeger {\emph{ et al.}} best.
Further, note that while the results reported by Jaeger {\emph{et al.}} refer
to the transducer response measured in ${\rm mV}$, our simulation results 
are normalized to a peak pressure height of unity. 

\section{Comparison of $3$D simulations to effectively $1$D approaches}
\label{sec:appendixB}

\begin{figure}[t!]
\centerline{\includegraphics{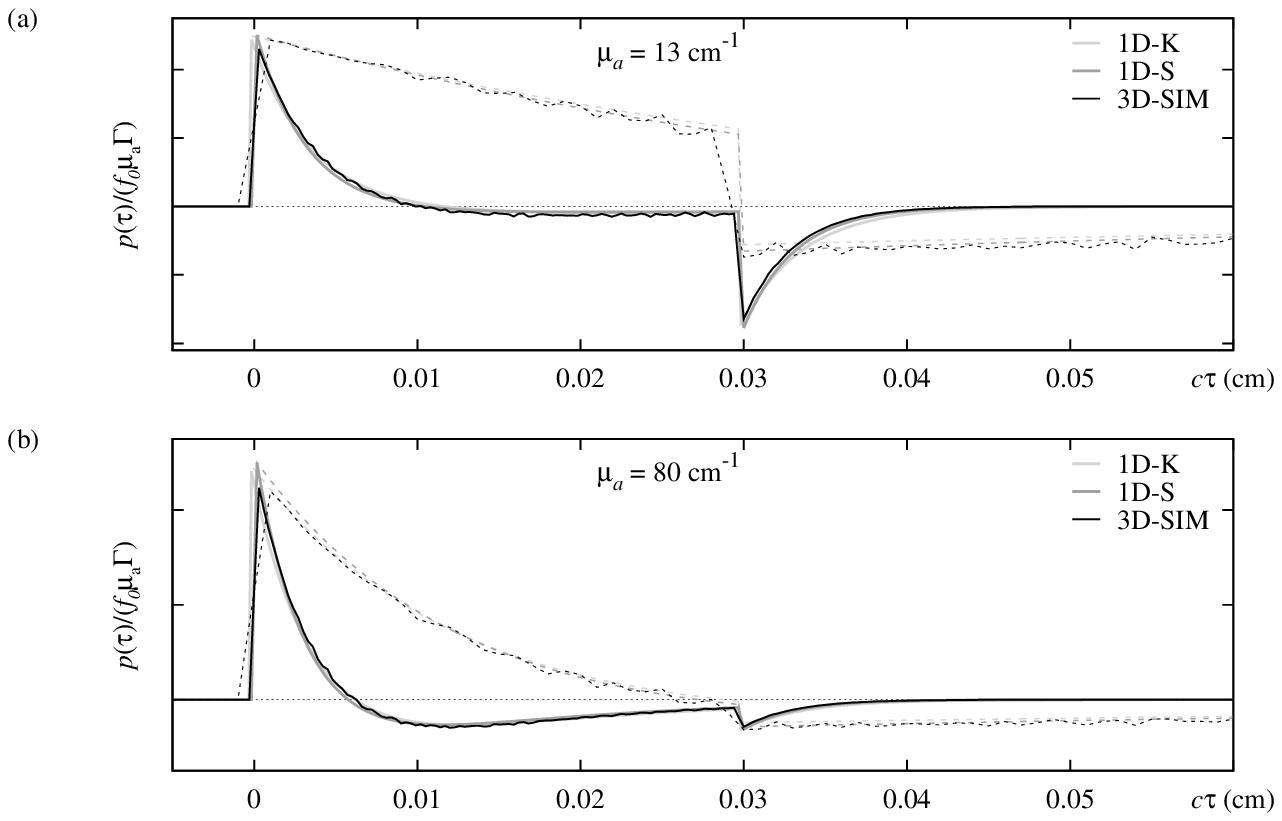}} 
\caption{Comparison of fully voxelized $3$D simulations using a Gaussian beam
profile (labeled ``$3$D-SIM''), following the simulation protocol discussed in sec.\
\ref{sec:theory}, to $1$D approaches that model the effect of 
acoustic diffraction under the assumption of plane acoustic waves and a 
Gaussian beam profile, see discussion in \ref{sec:appendixB}.
The curves labeled ``$1$D-K'' and ``$1$D-S'' result from numerical simulations
using the theoretical frameworks discussed by Ref.\ \cite{Karabutov:1996}
and Refs.\ \cite{Terzic:1984,Sigrist:1986}, respectively.
The source volume configuration is similar to the one discussed in terms of
numerical experiment $1$ in \ref{sec:appendixA}, i.e.\ (a) $\mu_a=13\,{\rm
cm}^{-1}$, and (b) $\mu_a=80\, {\rm cm}^{-1}$. 
In both subfigures solid (dashed) lines 
refer to the farfield (nearfield) scenario considering a Gaussian ISP with 
$1/e$-intensity threshold of $a_0=1.5\,{\rm mm}$ ($a_0=0.3\,{\rm mm}$).}
\label{fig:appendixB}
\end{figure}

As pointed out in sec.\ \ref{sec:theory}, we follow a numerical approach that
might be summarized as a three-step procedure: (i) modelling the optical
properties of a $3$D computational domain, (ii) considering the absorption and
scattering of laser light within the source volume, and, (iii) solving the
stress wave propagation problem for a medium with homogeneous acoustic
properties via the OA Poisson integral Eq.\ (\ref{eq:PAPoissonEq}).  While this
is a quite versatile $3$D approach that allows to model custom beam profiles
and intricate absorbing structures, note that there exist also effectively $1$D
approaches that yield OA waveforms under certain simplifying assumptions.
Here we consider two $1$D theoretical approaches, both based on the solution of
convolution integrals and valid under the assumption of plane acoustic
waves and Gaussian beam profiles.  Albeit these simplified approaches allow to
obtain closed form analytic OA signals, we here solve the respective equations
numerically for the source volume configuration discussed in terms of numerical
experiment $1$ in \ref{sec:appendixA}.  These simplified calculations serve as
limiting cases to which we compare our fully $3$D simulations, again allowing
for benchmarking our research code.  Below, both one-dimensional approaches are
briefly discussed. 

\paragraph{$1{\rm{D}}$ approach due to Karabutov {\emph{et al.}}:}
Considering the paraxial approximation of the $3$D acoustic wave equation and
assuming a transverse Gaussian ISP $f(\vec{r}_{\perp}) =
\exp(-|\vec{r}_{\perp}|^2/a_0^2)$, the diffraction transformation of OA signals
at the detection point $z$ along the beam axis (i.e.\ $\vec{r}_{\perp}=0$) can
be can be related to the unperturbed on-axis pressure profile $p_{\rm 0}(\tau)$
(wherein $\tau=t - z/c$)
via a linear convolution integral, \mbox{reading \cite{Karabutov:1996}}
\begin{equation}
p(\tau) = p_{\rm 0}(\tau) - \omega_{\rm D} \int_{-\infty}^{\tau}
        \!\mathsf{K}(\tau-\tau^\prime)\,p_{\rm 0}(\tau^\prime)
        \,\mathrm{d}\tau^\prime.
        \label{eq:OAV2}
\end{equation}
Therein the convolution kernel
\mbox{$\mathsf{K}(\tau-\tau^\prime)=\exp\{-\omega_{a} z L_{\rm
D}^{-1}(\tau-\tau^\prime)\}$} governs the diffraction transformation of the
propagating acoustic stress waves, depending on the characteristic frequency
$\omega_{a}=c \mu_a$ of the OA signal spectrum and the diffraction length
$L_{\rm D}=\pi a_0^2/\lambda$, defined by the acoustic wavelength 
$\lambda=2 \pi / \mu_a$.  Using the normalized initial stress
profile $p_{\rm 0}(\tau)=\exp\{-\mu_a c \tau\}$, nonzero in the range $c\tau
\in [0\,{\rm mm},0.3\,{\rm mm}]$, and the parameters $c=1.5\,{\rm mm/\mu s}$ as
well as $\mu_a$ and $a_0$ as reported by Ref.\ \cite{Paltauf:2000}, cf.\
numerical experiment $1$ in \ref{sec:appendixA}, we yield the curves labeled
by ``$1$D-K'' in \mbox{Fig.\ \ref{fig:appendixB}}.

\paragraph{$1{\rm{D}}$ approach due to Sigrist {\emph{et al.}}:}
A quite similar one-dimensional theoretical approach, build upon the assumption
of plane acoustic wave propagation, yields the convolution integral
\cite{Sigrist:1986,Terzic:1984}
\begin{equation}
p_r(\tau) = \omega_a \int_{-\infty}^{\infty}
        \!\mathsf{K}(\tau-\tau^\prime)\,p_{\rm 0}(\tau^\prime)
        \,\mathrm{d}\tau^\prime,
        \label{eq:OAV1}
\end{equation}
by means of which the diffraction altered 
excess pressure profile $p_r(\tau)$ for a rigid (i.e.\
motionless) boundary can be related to the unperturbed stress profile 
$p_{\rm 0}(\tau)$.
In our case, the more interesting excess pressure profile $p(\tau)$ for a free
(i.e.\ pressure-release) boundary is related to $p_r(\tau)$ simply via
$p(\tau) = \omega_a^{-1} \frac{d}{d\tau}p_r(\tau)$.
While the convolution kernel ${\mathsf{K}}$ is formally similar to the one
introduced previously, the definition of the diffraction length is slightly
different, i.e.\ $L_{\rm D} = 2 \pi d^2/\lambda^\prime$. Now, so as to treat
both one-dimensional frameworks under similar conditions, we effectively match
parameters by considering the acoustic wavelength used in Ref.\
\cite{Karabutov:1996}, i.e.\ $\lambda^\prime\equiv 2 \pi/ \mu_a$, and setting
the beam parameter $d\equiv a_0/\sqrt{2}$. The latter is in agreement with the
definition of $d$ as the diameter of the (Gaussian) beam spot with radial
profile $f^\prime(\vec{r}_{\perp}) = \exp(-2 |\vec{r}_{\perp}|^2/d^2)$, see
\cite{Sigrist:1986}.  Using the initial stress profile $p_{\rm 0}(\tau)$ and
simulation parameters $c$, $\mu_a$ and $a_0$ as above, we yield the curves
labeled by ``$1$D-S'' in Fig.\ \ref{fig:appendixB}.

First, note that the numerical results of the fully $3$D simulations via a
Gaussian beam profile agree very well with the effectively $1$D approaches
based on the additional assumption of plane acoustic wave propagation.  The
latter assumption in principle requires a setup in which $\mu_a a_0 \gg 1$, a
condition best satisfied for Fig.\ \ref{fig:appendixB}(b). The deviation of 
the data curves close to the steep increase and decrease of the pressure 
profile are solely due to the coarse mesh structure used in the $3$D approach.
Further, note that the simplified calculations in terms of purely Gaussian beam
profiles reproduce the compression and rarefaction characteristics of the
farfield signals seen in Figs.\ 8(a,b) of Ref.\ \cite{Paltauf:2000} quite well,
see the solid lines in Figs.\ \ref{fig:appendixB}(a,b). However, they cannot
account for the characteristic shape of the extended rarefaction phase caused
by the lateral limits of the beam profile observed for the nearfield
signals in Ref.\ \cite{Paltauf:2000}, see the dashed lines in
Figs.\ \ref{fig:appendixB}(a,b).
In contrast, note that the fully $3$D simulations using a custom flat-top
beam profile reported in \ref{sec:appendixA} reproduce the experiments 
of Ref.\ \cite{Paltauf:2000} satisfactorily.

\section*{References}
\bibliography{masterBibfile_OASignals,commentsBibfile_OASignals}

\end{document}